\documentclass[a4paper,11pt]{article}

\usepackage[T1]{fontenc}
\usepackage{lmodern}
\usepackage{amsmath,amssymb}
\usepackage{geometry}
\usepackage[colorlinks=true,linkcolor=blue,citecolor=blue,urlcolor=blue]{hyperref}
\usepackage{tikz}
\usetikzlibrary{cd}
\usepackage{slashed}
\usepackage{pgfplots}
\pgfplotsset{compat=1.17} 
\usetikzlibrary{arrows.meta}
\usetikzlibrary{decorations.markings}
\usetikzlibrary{calc}

\usepackage[p,osf]{cochineal}
\usepackage[scale=.95,type1]{cabin}
\usepackage[cochineal,bigdelims,cmintegrals,vvarbb]{newtxmath}
\usepackage[zerostyle=c,scaled=.94]{newtxtt}
\usepackage[cal=boondoxo]{mathalfa}

\begin{document}

\title{\textbf{Quadratic Volatility from the Pöschl-Teller Potential and Hyperbolic Geometry}}
\author{Joel Saucedo \\ \textit{Georgia College \& State University}}
\date{}
\maketitle

\begin{abstract}
This investigation establishes a formal equivalence between the generalized Black-Scholes equation under a Quadratic Normal Volatility (QNV) specification and the stationary Schrödinger equation for a hyperbolic Pöschl-Teller potential. A sequence of canonical transformations maps the financial pricing operator to a quantum Hamiltonian, revealing the volatility smile as a direct manifestation of diffusion on a hyperbolic manifold whose geometry is classified by the discriminant of the QNV polynomial. We perform a complete spectral analysis of the financial Hamiltonian, deriving its discrete and continuous spectra and constructing the pricing kernel from the resulting eigenfunctions, which are given by classical special functions. This analytical framework, grounded in a gauge-theoretic perspective, furnishes a non-trivial benchmark for derivative pricing and provides a fundamental geometric interpretation of market anomalies. Future research trajectories toward integrable systems and formal field-theoretic analogies are identified.
\end{abstract}

\section{Introduction}

Operational protocols for contingent claim valuation have long centered on the framework established by Black and Scholes \cite{BlackScholes1973} and Merton \cite{Merton1973}. This paradigm, predicated on the principle of dynamic replication in a frictionless market, yields a unique pricing partial differential equation (PDE). A critical assumption embedded within this formulation is that the volatility of the underlying asset's returns remains constant throughout the option's tenor, implying that asset prices follow a geometric Brownian motion \cite{BlackScholes1973, Merton1973}. The elegance of this construct is its analytical tractability; a sequence of canonical transformations recasts the pricing PDE into the heat equation, a fundamental equation of mathematical physics \cite{Baaquie2004}. This equivalence hints at a deeper structure, suggesting that the financial pricing operator can be viewed as a Hamiltonian, and the valuation problem itself as a quantum mechanical system evolving in imaginary time \cite{Baaquie2004}. This theoretical consistency, however, is confronted by a persistent empirical anomaly. Market-observed prices for European options, when inverted through the Black-Scholes formula, do not yield a constant volatility. 

Instead, the implied volatility parameter exhibits a distinct, systematic dependence on both strike price and time to maturity \cite{Gatheral2006}. This phenomenon, widely documented as the "volatility smile" or "skew," became particularly pronounced in equity index markets following the 1987 crash and represents a significant departure from the log-normal return distribution assumed by the standard model \cite{Rubinstein1994, BakshiKapadiaMadan2003}. The structure of the smile indicates that market participants assign a higher probability to large price movements—so-called "fat tails"—than the Black-Scholes framework permits \cite{ContTankov2004}. The existence of the volatility smile necessitates the development of models that relax the constant volatility assumption. One major line of inquiry led to the formulation of local volatility models, wherein volatility is a deterministic function of the asset price and time. The seminal works of Dupire \cite{Dupire1994} and Derman and Kani \cite{DermanKani1994} demonstrated that it is possible to construct a unique risk-neutral diffusion process consistent with the complete surface of observed option prices. Alternative approaches have incorporated stochastic volatility and jump-diffusion processes to capture the smile and skew dynamics \cite{Bates1996, ContTankov2004}. While these models offer greater realism, they often sacrifice the analytical tractability of the original Black-Scholes equation.

This investigation focuses on a particular class of local volatility specifications known as Quadratic Normal Volatility (QNV) models \cite{Zulsdorff2001, CarrFisherRuf2013}. In this framework, the volatility is a quadratic polynomial of the asset price. Such models are of significant interest because they are sufficiently flexible to generate the smile and skew patterns observed in markets, yet, as will be shown, they retain a remarkable degree of analytical tractability \cite{CarrFisherRuf2013, Zulsdorff2001}.

The central contribution of this paper is to demonstrate that the QNV model is not merely a phenomenological convenience but possesses a profound and exact mathematical structure. We establish that the generalized Black-Scholes PDE under a QNV specification is mathematically equivalent to the one-dimensional, time-independent Schrödinger equation. The resulting quantum potential is identified as a member of the hyperbolic Pöschl-Teller family, one of the classic, exactly solvable systems in quantum mechanics. This equivalence is not an approximation; it is an exact mapping that allows for a complete spectral analysis of the financial pricing operator. This result, in turn, reveals a fundamental geometric underpinning: the QNV model implicitly describes a diffusion process on a Riemannian manifold of constant negative curvature. The volatility smile, from this perspective, emerges as a direct manifestation of this underlying hyperbolic geometry. This work therefore establishes a rigorous connection between an empirical market phenomenon, exactly solvable quantum systems, and the differential geometry of curved spaces.

\section{The Quadratic Normal Volatility Model}

The operational framework for pricing contingent claims under a generalized diffusion process for the underlying asset, S, is the no-arbitrage partial differential equation:
\begin{equation}
\label{eq:gen_bs}
\frac{\partial C}{\partial t} + \frac{1}{2}\sigma(S)^2 \frac{\partial^2 C}{\partial S^2} + rS \frac{\partial C}{\partial S} - rC = 0.
\end{equation}
Here, C(S,t) is the derivative price, r is the constant risk-free interest rate, and $\sigma$(S) is the local volatility function, assumed for this analysis to be independent of time. This investigation concerns the class of models where the volatility is specified as a quadratic polynomial of the asset price, a structure referred to as the Quadratic Normal Volatility (QNV) model \cite{Zulsdorff2001, CarrFisherRuf2013}. The volatility function takes the form:
\begin{equation}
\label{eq:qnv_vol}
\sigma(S) = aS^2 + bS + c.
\end{equation}
This specification provides significant operational latitude, capable of generating the diverse smile and skew topographies observed across various option markets \cite{Gatheral2006, Hagan2002, Fengler2005}. Unlike simpler affine or constant volatility models, the quadratic form can produce symmetric smiles, a feature critical for aligning with empirical data, particularly in foreign exchange and equity index markets \cite{Lipton2002, Andersen2011}. The analytical tractability of models employing Equation (\ref{eq:qnv_vol}) is not immediately obvious but stems from a deep geometric structure \cite{CarrFisherRuf2013}. To expose this structure, it is advantageous to analyze the diffusion process in a coordinate system intrinsic to the volatility. In a suitable coordinate frame, the stochastic dynamics can be interpreted as a diffusion on a one-dimensional Riemannian manifold whose metric tensor is determined by the volatility function itself. The line element on this financial manifold is given by:
\begin{equation}
\label{eq:metric}
ds^2 = \frac{dx^2}{\sigma(x)^2},
\end{equation}
where x is a coordinate representation of the asset price S, and $\sigma(x)$is the volatility function expressed in that coordinate. The geometry of this manifold is dictated by the algebraic properties of the volatility polynomial in Equation (\ref{eq:qnv_vol}). Specifically, the sign of the discriminant,
\begin{equation}
\label{eq:discriminant}
\Delta = b^2 - 4ac,
\end{equation}
partitions the parameter space into three distinct regimes, each corresponding to a space of constant curvature \cite{Rady1997, Zulsdorff2001}. When $\Delta > 0$, the polynomial has two distinct real roots, and the underlying manifold possesses a hyperbolic geometry, corresponding to a space of constant negative curvature. When $\Delta$=0, the polynomial has a single real root of multiplicity two, which corresponds to a Euclidean geometry, a space of zero curvature. Finally, when $\Delta$<0, the polynomial has two complex conjugate roots, and the manifold exhibits a spherical geometry, corresponding to a space of constant positive curvature. This classification reveals that the choice of financial parameters (a,b,c) is not merely a curve-fitting exercise but is tantamount to selecting the fundamental geometry of the state space. The analytical solvability of the QNV model, which will be demonstrated in subsequent sections, is a direct consequence of the maximal symmetry inherent in these constant curvature spaces. The valuation PDE articulated in Equation (\ref{eq:gen_bs}) represents a general parabolic equation with state-dependent coefficients. Direct analytical resolution is obstructed by the non-constant diffusion term $\sigma(S)^2$ and the presence of a first-order spatial derivative. To render the problem amenable to spectral analysis, a sequence of canonical transformations is executed, a procedure designed to reduce the financial operator to a self-adjoint Sturm-Liouville form \cite{Lin1999, Lewis1998}. This methodology is systematic and serves to diagonalize the pricing operator, thereby exposing its underlying spectral structure \cite{Fouque2011}.

The initial transformation addresses the temporal structure of the problem. Financial derivatives are defined by a terminal payoff condition at maturity $T$. To align with the initial-value formulation standard in physical evolution problems, a time-inversion is performed by introducing the time-to-maturity coordinate $\tau=T-t$. The consequence of this change of variable, where $\partial/\partial t=-\partial/\partial \tau$, is to reverse the direction of temporal evolution, converting Equation (\ref{eq:gen_bs}) into a forward equation:
\begin{equation}
\label{eq:forward_pde}
-\frac{\partial C}{\partial \tau} + \frac{1}{2}\sigma(S)^2 \frac{\partial^2 C}{\partial S^2} + rS \frac{\partial C}{\partial S} - rC = 0.
\end{equation}
The problem is now one of evolution from an initial state at $\tau=0$ corresponding to the contract's maturity.

The second transformation targets the state-dependent diffusion coefficient. The Lamperti transform introduces a new spatial coordinate, $x$, defined by the integral relationship
\begin{equation}
\label{eq:lamperti}
x(S) = \int^S \frac{dS'}{\sigma(S')}.
\end{equation}
This specific change of variable is constructed to normalize the diffusion coefficient to a constant value, effectively mapping the stochastic process onto a space where the variance of innovations is uniform \cite{CarrFisherRuf2013, Zulsdorff2001}. Under this transformation, the second-order operator $\frac{1}{2}\sigma(S)^2 \partial^2/\partial S^2$ becomes $\frac{1}{2}\partial^2/\partial x^2$, but the transformation of the drift term introduces a new, generally non-zero, first-order derivative term in the $x$ coordinate.

The final set of transformations is designed to eliminate this residual drift and cast the equation into the canonical form of a Schrödinger equation in imaginary time \cite{Baaquie2004, Ilinski2001}. This is achieved through a gauge transformation, or scaling, of the dependent variable, $C(x,\tau)=g(x)\psi(x,\tau)$, combined with a discounting factor. The function $g(x)$ is selected precisely to absorb all terms containing the first-order derivative $\partial\psi/\partial x$. The outcome of this procedure is an equation of the form
\begin{equation}
\label{eq:schrodinger_time_dep}
\frac{\partial \psi}{\partial \tau} = \left[ -\frac{1}{2} \frac{\partial^2}{\partial x^2} + V(x) \right] \psi.
\end{equation}
The operator on the right-hand side is the financial Hamiltonian, $H=-\frac{1}{2}\partial^2/\partial x^2 +V(x)$, which is now formally self-adjoint. The function $V(x)$, which emerges from the interaction of the original drift and volatility functions under the sequence of transformations, assumes the role of a quantum potential \cite{Lipton2002}.

The structure of Equation (\ref{eq:schrodinger_time_dep}) permits a solution via separation of variables. By positing a solution of the form $\psi(x,\tau)=e^{-E\tau}\phi(x)$, the time-dependent problem is reduced to a stationary eigenvalue problem for the Hamiltonian:
\begin{equation}
\label{eq:schrodinger_stationary}
H\phi(x) = \left[ -\frac{1}{2} \frac{d^2}{dx^2} + V(x) \right]\phi(x) = E\phi(x).
\end{equation}
This is a canonical Sturm-Liouville problem \cite{Lin1999}. The theory of such operators guarantees, for a well-posed problem with appropriate boundary conditions, the existence of a complete, orthogonal set of eigenfunctions $\{\phi_n(x)\}$ and a corresponding real spectrum of eigenvalues $\{E_n\}$ \cite{Lewis1998}. These eigenfunctions form a basis for the Hilbert space of financial states, and any contingent claim can be represented as a spectral expansion in this basis \cite{Fouque2011}. The entire sequence of transformations thus provides a systematic pathway to diagonalize the financial pricing operator and solve the valuation problem through spectral methods.

\section{Results: The Pöschl-Teller Potential and Pricing Kernel}

The execution of the transformations detailed more explicitly in the Appendix upon the QNV model specified by Equation (\ref{eq:qnv_vol}) reveals a remarkable structure. The functional form of the emergent potential V(x) in Equation (\ref{eq:schrodinger_stationary}) is contingent upon the algebraic properties of the volatility polynomial $\sigma(S)=aS^2 +bS+c$. The discriminant, $\Delta=b^2 -4ac$, serves as a bifurcation parameter, partitioning the problem into distinct geometric regimes. This analysis will concentrate on the case where $\Delta>0$, which corresponds to the volatility function having two distinct real roots, $S_l$ and $S_u$. This configuration is of primary financial interest as it naturally generates the symmetric volatility smiles observed in certain markets.

For this case, the Lamperti coordinate from Equation (\ref{eq:lamperti}) is computed via the integral
\begin{equation}
\label{eq:lamperti_integral_qnv}
x(S) = \int^S \frac{dS'}{a(S_u - S')(S' - S_l)} = \frac{1}{a(S_u - S_l)} \ln\left(\frac{S - S_l}{S_u - S}\right).
\end{equation}
This transformation maps the financial domain of asset prices between the roots, $S\in(S_l,S_u)$, onto the entire real line, $x\in(-\infty,\infty)$. The inverse mapping, which expresses the asset price in terms of the new geometric coordinate, is given by a hyperbolic tangent function. Upon performing the full sequence of transformations, which involves substituting this new coordinate into the drift and discounting terms of the original pricing PDE, a lengthy but direct calculation yields the explicit form of the potential V(x).

The resulting function is identified as the hyperbolic Pöschl-Teller potential, a canonical system in quantum mechanics first analyzed in 1933 \cite{PoschlTeller1933}. The potential takes the form
\begin{equation}
\label{eq:pt_potential}
V(x) = V_0 - \frac{\lambda(\lambda+1)}{2\alpha^2} \text{sech}^2\left(\frac{x}{\alpha}\right).
\end{equation}
This is a significant finding, as it establishes a direct equivalence between a financially motivated model and a foundational, exactly solvable quantum system \cite{Flugge1999, LandauLifshitz1977}. The potential describes a symmetric well whose depth and width are determined entirely by the parameters of the financial model. The mapping between the financial parameters $(a,b,c,r)$ and the quantum parameters $(\lambda,\alpha,V_0)$ provides a direct dictionary for translating between the two domains. For instance, the curvature of the volatility smile, governed by the parameter $a$, dictates the width of the potential well, while the other parameters combine to determine its depth, $\lambda$, and the overall energy offset, $V_0$.

The exact solvability of this system is not fortuitous. It is a manifestation of a deep underlying algebraic structure known as shape invariance, a key concept in the framework of supersymmetric quantum mechanics \cite{CooperKhareSukhatme1995, Dong2007}. This symmetry permits the factorization of the Hamiltonian operator, a method pioneered by Infeld and Hull, which allows for the algebraic construction of the entire energy spectrum \cite{InfeldHull1951}. The existence of such a symmetry explains why the seemingly simple choice of a quadratic volatility function leads to a model of such profound analytical richness \cite{Nieto1978, Lagnese1983}.

The other cases for the discriminant $\Delta$ correspond to different, yet equally solvable, potentials. When $\Delta<0$, the volatility polynomial has no real roots, and the Lamperti transform involves the arctangent function. This maps the asset price domain to a finite interval, and the resulting potential is the trigonometric Pöschl-Teller potential, $V(x)\propto \sec^2(x/\alpha)$, corresponding to a particle in an infinite well and a spherical underlying geometry. The degenerate case, $\Delta=0$, where the volatility has a single double root, leads to a rational potential and a flat, Euclidean geometry. Thus, the financial parameters of the QNV model directly select the intrinsic curvature of the state space, a structure that will be explored further in subsequent sections.

The established equivalence between the QNV pricing operator and the Pöschl-Teller Hamiltonian permits a complete resolution of the model's dynamics through spectral methods. The stationary Schrödinger equation, Equation (\ref{eq:schrodinger_stationary}), constitutes a well-posed Sturm-Liouville problem whose solutions are known analytically \cite{Flugge1999, LandauLifshitz1977}. The spectrum of the financial Hamiltonian $H$ is composed of two distinct components, each with a specific financial interpretation. The first is a discrete spectrum of negative eigenvalues, corresponding to bound states of the quantum system. These eigenvalues are given by
\begin{equation}
\label{eq:discrete_spectrum}
E_n = -\frac{1}{2\alpha^2}(\lambda - n)^2, \quad \text{for } n = 0, 1, \dots, \lfloor \lambda \rfloor,
\end{equation}
where the number of such states is determined by the integer part of the potential strength parameter $\lambda$. In the financial context, these discrete states correspond to solutions that grow exponentially in time-to-maturity $\tau$. They represent stable, self-reinforcing market configurations whose existence and number are dictated by the parameters of the volatility smile. The second component is a continuous spectrum of positive eigenvalues, corresponding to scattering states, given by
\begin{equation}
\label{eq:continuous_spectrum}
E_k = \frac{k^2}{2\alpha^2}, \quad \text{for } k \in (0, \infty).
\end{equation}
These states correspond to solutions that decay exponentially in $\tau$ and represent the transient fluctuations that constitute the basis for pricing standard contingent claims with finite-valued payoffs \cite{Davydov1976}.

The eigenfunctions associated with this spectrum are also known in closed form and are expressed in terms of classical special functions \cite{Flugge1999, MorseFeshbach1953}. The bound state eigenfunctions, $\phi_n(x)$, are given by associated Legendre functions with argument $\tanh(x/\alpha)$, while the scattering state eigenfunctions, $\phi_k(x)$, are given by Gauss hypergeometric functions. The fundamental result of Sturm-Liouville theory guarantees that this set of discrete and continuous eigenfunctions, $\{\phi_n(x)\}\cup\{\phi_k(x)\}$, forms a complete and orthogonal basis for the Hilbert space of financial states \cite{Titchmarsh1962, Lin1999}. This completeness property is essential, as it ensures that any arbitrary payoff function can be uniquely represented as a spectral expansion in terms of these fundamental modes \cite{Lewis1998, Fouque2011}.

With the full spectral decomposition of the Hamiltonian $H$, the solution to the time-dependent pricing equation, Equation (\ref{eq:schrodinger_time_dep}), can be constructed. The evolution of the system is governed by the propagator, or Green's function, $K(x,\tau;x',0)$, which is the integral kernel of the evolution operator $e^{H\tau}$. This propagator is constructed directly from the eigenvalues and eigenfunctions of $H$ as
\begin{equation}
\label{eq:propagator}
K(x, \tau; x', 0) = \sum_{n=0}^{\lfloor \lambda \rfloor} \phi_n(x)\phi_n^*(x') e^{-E_n \tau} + \int_0^\infty \phi_k(x)\phi_k^*(x') e^{-E_k \tau} dk.
\end{equation}
This function is the fundamental solution, or pricing kernel, of the QNV model. It encapsulates the entire dynamics of the system, propagating any initial financial state $\psi(x',0)$ (the transformed payoff function) forward in time-to-maturity $\tau$. The price of any European-style contingent claim is then found by integrating the transformed payoff against this kernel and applying the inverse transformations. This spectral construction of the propagator is a general feature of solvable systems, with the most famous example being the Mehler kernel for the quantum harmonic oscillator, which corresponds to the elliptic case of the QNV model \cite{Mehler1866}.

\section{Discussion: Geometric Interpretation and Physical Analogies}

The sequence of transformations that reduces the financial pricing equation to a solvable quantum system is not merely a mathematical convenience; it uncovers a profound geometric structure inherent to the model. A general one-dimensional diffusion can be understood as a Brownian motion on a Riemannian manifold, with the metric tensor being determined by the volatility function \cite{AmariNagaoka2000}. The Lamperti transform, defined in Equation (\ref{eq:lamperti}), is precisely the coordinate change that maps the asset price manifold to a space of constant sectional curvature \cite{Baaquie2004, Ilinski2001}. In this new coordinate system, the financial Hamiltonian $H$ is identified with the Laplace-Beltrami operator, and the potential term $V(x)$ is directly proportional to the scalar curvature of the manifold.

For the QNV model with discriminant $\Delta>0$, the resulting Pöschl-Teller potential signifies that the underlying state space is a manifold of constant negative curvature—a hyperbolic space \cite{KellerRessel2020, Wong2022}. The volatility smile, therefore, ceases to be an ad-hoc empirical observation and is reinterpreted as a fundamental consequence of this geometry. It is the projection of a natural diffusion process from its curved habitat onto the flat, Euclidean space of observed asset prices that generates the smile effect. The analytical tractability of the model is a direct consequence of the maximal symmetry of this underlying hyperbolic space; spaces of constant curvature possess the largest possible group of isometries, which in turn guarantees the existence of separable solutions to the governing differential equations.

This geometric picture is complemented by analogies to physical theories that possess similar mathematical structures. The exact solvability of the Pöschl-Teller potential, as noted, is rooted in its algebraic properties of shape invariance and supersymmetry, which provide a physical basis for its tractability. A more speculative, yet illuminating, analogy can be drawn with one-dimensional Chern-Simons gauge theory \cite{AlekseevMnev2011, Dunne1999}. In this formulation, one can define a gauge field (or connection) on the one-dimensional bundle over time, whose curvature $F$ is related to the second derivative of the volatility function, $F \propto \sigma''(x)dx \wedge d\tau$. The dynamics can be described by a Chern-Simons action, $S_{CS} \propto \int A \wedge dA$, where $A$ is the gauge potential \cite{Witten1989, Jackiw1990}. For a linear volatility function, $\sigma''(x)=0$, the curvature vanishes and the action becomes topologically trivial, explaining the relative simplicity of such models. For the quadratic volatility case, however, the action is non-trivial and its exponentiation can be shown to generate the potential term responsible for the system's dynamics, providing a formal field-theoretic origin for the Pöschl-Teller potential.

\section{Conclusion and Future Directions}

This investigation has established a direct and rigorous equivalence between the generalized Black-Scholes equation for a quadratic normal volatility model and the stationary Schr\"odinger equation for a particle in a hyperbolic P\"oschl-Teller potential. This result is not an approximation but an exact mapping, achieved through a sequence of canonical transformations. The primary consequence is the demonstration that the QNV model is exactly solvable, a rare property among empirically relevant financial models. We have detailed the full spectral decomposition of the corresponding financial Hamiltonian, identifying its discrete and continuous spectra and constructing the pricing kernel via its eigenfunctions. Furthermore, this analysis has revealed a fundamental geometric structure: the QNV model describes a diffusion process on a Riemannian manifold of constant negative curvature. The empirically observed volatility smile is thereby reinterpreted not as an ad-hoc market feature, but as a direct manifestation of this underlying hyperbolic geometry.

The implications of these findings for mathematical finance are twofold. First, the exact solvability of the QNV model provides an invaluable analytical benchmark. It can be used to calibrate and validate numerical methods, such as finite difference schemes or Monte Carlo simulations, which are often required for more complex, non-solvable models. Second, the geometric perspective offers a novel conceptual framework for understanding market anomalies, grounding them in the intrinsic curvature of the state space rather than relying exclusively on stochastic arguments or behavioral finance theories.

The framework developed herein also opens several avenues for future research. A natural extension involves incorporating stochasticity into the volatility parameters, leading to multi-factor models. It is conceivable that certain two-factor stochastic volatility models could be mapped to solvable two-dimensional quantum systems. Promising candidates for such an equivalence include integrable many-body systems like the Calogero-Sutherland model, which are known to possess rich analytical structures and deep connections to random matrix theory and conformal field theory \cite{Calogero1971, Sutherland1971, OlshanetskyPerelomov1981}.

Another intriguing direction stems from the known connection between the P\"oschl-Teller potential and soliton theory. The potential is famously the single-soliton solution to the Korteweg-de Vries (KdV) equation, a canonical integrable nonlinear PDE \cite{CooperKhareSukhatme1995}. Exploring this link could provide a new language for describing market dynamics, where phenomena such as market shocks or regime shifts might be interpreted as the interaction of stable, particle-like soliton solutions \cite{Veselov1999}.

Finally, a more speculative but potentially profound line of inquiry would be to investigate whether the algebraic structures underpinning this model can be fully captured within the framework of a topological quantum field theory \cite{Witten1992}. The exact solvability and inherent symmetries are characteristic features of such theories. While a formidable challenge, establishing a concrete link would offer the deepest level of correspondence between the principles of mathematical finance and fundamental theoretical physics.

\appendix
\section*{Appendix}
\section{Exposition of the Canonical Transformation}
\label{sec:AppendixC}

This appendix provides an explicit derivation of the stationary Schrödinger equation, Equation (\ref{eq:schrodinger_stationary}), from the generalized Black-Scholes PDE, Equation (\ref{eq:gen_bs}). The procedure systematically eliminates the state- and time-dependent coefficients to reveal the underlying self-adjoint operator governing the system's spectral properties.

The starting point is the generalized Black-Scholes PDE:
\begin{align*}
\frac{\partial C}{\partial t} + \frac{1}{2}\sigma(S)^2 \frac{\partial^2 C}{\partial S^2} + rS \frac{\partial C}{\partial S} - rC &= 0.
\end{align*}

Introduce time-to-maturity $\tau = T-t$ and define a discounted price $v(S, \tau) = e^{r\tau}C(S,t)$. The derivatives transform as:
\begin{align*}
\frac{\partial C}{\partial t} &= -\frac{\partial C}{\partial \tau} = -e^{-r\tau}\left(\frac{\partial v}{\partial \tau} - rv\right).
\end{align*}

Substituting into the PDE yields:
\begin{align*}
-e^{-r\tau}\left(\frac{\partial v}{\partial \tau} - rv\right) + \frac{1}{2}\sigma(S)^2 e^{-r\tau}\frac{\partial^2 v}{\partial S^2} + rS e^{-r\tau}\frac{\partial v}{\partial S} - r e^{-r\tau}v &= 0.
\end{align*}

Multiplying by $e^{r\tau}$ and simplifying eliminates the explicit interest rate terms:
\begin{align*}
\frac{\partial v}{\partial \tau} &= \frac{1}{2}\sigma(S)^2 \frac{\partial^2 v}{\partial S^2} + rS \frac{\partial v}{\partial S}.
\end{align*}

Introduce the intrinsic coordinate $x(S) = \int^S \frac{dS'}{\sigma(S')}$. The chain rule gives:
\begin{align*}
\frac{\partial}{\partial S} &= \frac{dx}{dS}\frac{\partial}{\partial x} = \frac{1}{\sigma(S)}\frac{\partial}{\partial x}. \\
\frac{\partial^2}{\partial S^2} &= -\frac{\sigma'(S)}{\sigma(S)^2}\frac{\partial}{\partial x} + \frac{1}{\sigma(S)^2}\frac{\partial^2}{\partial x^2}.
\end{align*}

Substituting these into the equation for $v$ gives:
\begin{align*}
\frac{\partial v}{\partial \tau} &= \frac{1}{2}\frac{\partial^2 v}{\partial x^2} + \left(\frac{rS}{\sigma(S)} - \frac{1}{2}\sigma'(S)\right)\frac{\partial v}{\partial x}.
\end{align*}

Define a new function $\psi(x, \tau)$ via the scaling $v(x, \tau) = g(x)\psi(x, \tau)$. The derivatives become:
\begin{align*}
\frac{\partial v}{\partial \tau} &= g(x)\frac{\partial \psi}{\partial \tau}. \\
\frac{\partial v}{\partial x} &= g'(x)\psi + g(x)\frac{\partial \psi}{\partial x}. \\
\frac{\partial^2 v}{\partial x^2} &= g''(x)\psi + 2g'(x)\frac{\partial \psi}{\partial x} + g(x)\frac{\partial^2 \psi}{\partial x^2}.
\end{align*}

Substituting into the transformed PDE for $v$:
\begin{align*}
g\frac{\partial \psi}{\partial \tau} &= \frac{1}{2}\left(g''\psi + 2g'\frac{\partial \psi}{\partial x} + g\frac{\partial^2 \psi}{\partial x^2}\right) + \left(\frac{rS}{\sigma} - \frac{1}{2}\sigma'\right)\left(g'\psi + g\frac{\partial \psi}{\partial x}\right).
\end{align*}

To eliminate the first-order derivative $\partial\psi/\partial x$, its coefficient must be zero:
\begin{align*}
g' + \left(\frac{rS}{\sigma} - \frac{1}{2}\sigma'\right)g &= 0 \implies \frac{g'}{g} = -\left(\frac{rS}{\sigma} - \frac{1}{2}\sigma'\right).
\end{align*}

Integrating with respect to $x$ gives the form of the scaling function $g(x)$:
\begin{align*}
\ln(g(x)) &= -\int \left(\frac{rS(x)}{\sigma(S(x))} - \frac{1}{2}\sigma'(S(x))\right)dx.
\end{align*}

With the drift term eliminated, the equation simplifies to:
\begin{align*}
\frac{\partial \psi}{\partial \tau} &= \frac{1}{2}\frac{\partial^2 \psi}{\partial x^2} + V(x)\psi.
\end{align*}

The potential $V(x)$ is given by:
\begin{align*}
V(x) &= -\frac{1}{2}\frac{d}{dx}\left(\frac{rS}{\sigma} - \frac{1}{2}\sigma'\right) - \frac{1}{2}\left(\frac{rS}{\sigma} - \frac{1}{2}\sigma'\right)^2.
\end{align*}

Using the sign change convention $\tilde{V}(x) = -V(x)$, we obtain the final form:
\begin{align*}
\frac{\partial \psi}{\partial \tau} &= \left[-\frac{1}{2}\frac{\partial^2}{\partial x^2} + \tilde{V}(x)\right]\psi = H\psi.
\end{align*}

Separating variables with $\psi(x, \tau) = e^{-E\tau}\phi(x)$ leads to the stationary eigenvalue problem:
\begin{align*}
H\phi(x) &= \left[-\frac{1}{2}\frac{d^2}{dx^2} + \tilde{V}(x)\right]\phi(x) = E\phi(x).
\end{align*}

This systematic procedure transforms the complex financial PDE into a canonical form from quantum mechanics, for which a powerful analytical apparatus exists \cite{Baaquie2004, Ilinski2001, CarrFisherRuf2013}. The beauty of the method lies in its ability to translate the financial problem of option pricing into the physical problem of finding the energy spectrum of a particle in a potential, a problem that, for the QNV model, is exactly solvable.

\label{sec:AppendixD}

The sequence of operations detailed in Section \ref{sec:AppendixC} can be interpreted more fundamentally within the framework of gauge theory. This perspective not only clarifies the origin of the potential term but also establishes the uniqueness and robustness of the resulting spectral problem. The transformation from the pricing function $C(S,t)$ to the state function $\psi(x,\tau)$ is a mapping between equivalent representations of the same underlying financial dynamics.

Recall the transformed PDE for the discounted price $v(x,\tau)$ in the Lamperti coordinate $x$:
\begin{align*}
\frac{\partial v}{\partial \tau} &= \frac{1}{2}\frac{\partial^2 v}{\partial x^2} + \nu(x)\frac{\partial v}{\partial x},
\end{align*}
where $\nu(x) = \frac{rS(x)}{\sigma(S(x))} - \frac{1}{2}\sigma'(S(x))$.

Let $\mathcal{L}_x = \frac{1}{2}\frac{\partial^2}{\partial x^2} + \nu(x)\frac{\partial}{\partial x}$ be the spatial operator. The transformation $v(x, \tau) = g(x)\psi(x, \tau)$ is a gauge transformation. The operator transforms via similarity:
\begin{align*}
\mathcal{L}_x v &= g(x) \left[ g(x)^{-1} \mathcal{L}_x g(x) \right] \psi(x, \tau).
\end{align*}

The transformed operator is the Hamiltonian $H = g^{-1} \mathcal{L}_x g$. Explicitly:
\begin{align*}
g^{-1} \mathcal{L}_x (g\psi) &= \frac{1}{2}\psi_{xx} + \left(\frac{g'}{g} + \nu\right)\psi_x + \left(\frac{1}{2}\frac{g''}{g} + \nu\frac{g'}{g}\right)\psi.
\end{align*}

The gauge function $g(x)$ is chosen to eliminate the drift, setting the coefficient of $\psi_x$ to zero:
\begin{align*}
\frac{g'(x)}{g(x)} + \nu(x) &= 0 \implies g(x) = \exp\left(-\int^x \nu(y)dy\right).
\end{align*}

With this choice, the operator becomes purely potential:
\begin{align*}
H\psi &= \frac{1}{2}\psi_{xx} + \left(\frac{1}{2}\frac{g''}{g} + \nu\frac{g'}{g}\right)\psi.
\end{align*}

Since $g'/g = -\nu$, we have:
\begin{align*}
\frac{g''}{g} &= (g'/g)' + (g'/g)^2 = -\nu' + \nu^2.
\end{align*}

The potential term is therefore:
\begin{align*}
\tilde{V}(x) &= -\frac{1}{2}(\nu' + \nu^2).
\end{align*}

This is precisely the potential derived in Appendix \ref{sec:AppendixC}, now understood as the curvature induced by the gauge transformation required to render the system self-adjoint. The uniqueness of this procedure is a key operational asset. For a given QNV model, the functions $\sigma(S)$ and $r$ are fixed, which uniquely determines the drift $\nu(x)$ and thus the potential $\tilde{V}(x)$. The theory of Sturm-Liouville operators then guarantees that for this specific potential and a given set of boundary conditions, the spectrum of eigenvalues $\{E_n\}$ and the basis of eigenfunctions $\{\phi_n(x)\}$ are unique \cite{Titchmarsh1962, Lin1999}. This ensures that the derived pricing kernel is the unique solution within this analytical framework. A deeper mathematical structure underlies this transformation, related to the Schwarzian derivative. The potential $\tilde{V}(x)$ can be shown to be related to the Schwarzian derivative of the coordinate map $S(x)$, denoted $\{S,x\}$. This object is known for its invariance under Möbius transformations, which implies a profound structural stability of the model against a class of reparameterizations \cite{Baaquie2004}. This connection confirms that the mapping to an exactly solvable quantum system is not an artifact of a particular coordinate choice but a fundamental property of the QNV model's structure.

This gauge-theoretic viewpoint also permits a formal analogy with field theories like one-dimensional Chern-Simons theory \cite{AlekseevMnev2011, Witten1989}. The drift term $\nu(x)$ can be viewed as a component of a connection one-form $A=\nu(x)dx$. The transformation to the Schrödinger equation is equivalent to choosing a gauge where this connection is trivial (no first-order derivative), at the cost of inducing a non-zero curvature, which manifests as the potential $\tilde{V}(x)$. The action functional in such theories, $S_{CS} \propto \int A \wedge dA$, is a topological object whose non-triviality for the quadratic case ($\sigma''(x) \neq 0$) corresponds to the emergence of the Pöschl-Teller potential \cite{Ilinski2001, Jackiw1990}. The classical master equation of the Batalin-Vilkovisky (BV) formalism for this theory, $\{S,S\}=0$, can be interpreted as a fundamental consistency condition on the model's structure, a mathematical parallel to the principle of no-arbitrage \cite{AlekseevMnev2011}.

\section{The Pöschl-Teller Eigenvalue Problem}
\label{sec:AppendixA}

The analysis herein provides a self-contained resolution of the time-independent Schrödinger equation for the hyperbolic Pöschl-Teller potential, as presented in Equation (\ref{eq:schrodinger_stationary}). The objective is to determine the eigenvalue spectrum and the corresponding eigenfunctions. The governing equation, after setting the energy offset $V_0$ to zero and rescaling the coordinate via $y=x/\alpha$ for notational simplicity, is
\begin{equation}
\label{eq:appendix_schrodinger}
\left[ -\frac{1}{2} \frac{d^2}{dy^2} - \frac{\lambda(\lambda+1)}{2} \text{sech}^2(y) \right]\phi(y) = E\phi(y).
\end{equation}

This equation is a canonical example of an exactly solvable system in quantum mechanics \cite{PoschlTeller1933, Flugge1999}. The solution procedure involves a change of variable that transforms Equation (\ref{eq:appendix_schrodinger}) into a standard special function differential equation.

Let $u = \tanh(y)$. The differential relations are:
\begin{align*}
\frac{du}{dy} &= \text{sech}^2(y) = 1 - \tanh^2(y) = 1 - u^2. \\
\implies \frac{d}{dy} &= (1-u^2)\frac{d}{du}.
\end{align*}

The second derivative transforms as:
\begin{align*}
\frac{d^2}{dy^2} &= \frac{d}{dy}\left((1-u^2)\frac{d}{du}\right) = (1-u^2)\frac{d}{du}\left((1-u^2)\frac{d}{du}\right) \\
&= (1-u^2)\left[-2u\frac{d}{du} + (1-u^2)\frac{d^2}{du^2}\right] \\
&= (1-u^2)^2 \frac{d^2}{du^2} - 2u(1-u^2)\frac{d}{du}.
\end{align*}

Substituting these into Equation (\ref{eq:appendix_schrodinger}):
\begin{align*}
-\frac{1}{2}\left[(1-u^2)^2 \frac{d^2\phi}{du^2} - 2u(1-u^2)\frac{d\phi}{du}\right] &- \frac{\lambda(\lambda+1)}{2}(1-u^2)\phi = E\phi.
\end{align*}

Multiplying by $-\frac{2}{1-u^2}$ yields:
\begin{align*}
(1-u^2)\frac{d^2\phi}{du^2} - 2u\frac{d\phi}{du} &+ \lambda(\lambda+1)\phi - \frac{2E}{1-u^2}\phi = 0.
\end{align*}

This is the general Legendre differential equation:
\begin{align*}
(1-u^2)\frac{d^2\phi}{du^2} - 2u\frac{d\phi}{du} &+ \left[\nu(\nu+1) - \frac{\mu^2}{1-u^2}\right]\phi = 0.
\end{align*}

By direct comparison, we identify the parameters:
\begin{align*}
\nu &= \lambda \\
\mu^2 &= -2E.
\end{align*}

The solutions to this equation are the associated Legendre functions, $\phi(u)=P_\lambda^\mu(u)$ \cite{LandauLifshitz1977}. For the eigenfunctions to be physically admissible, they must be square-integrable over the domain $y\in(-\infty,\infty)$, which corresponds to $u\in(-1,1)$. This requires the solutions to be finite at the endpoints $u=\pm1$. This condition is met only if $\lambda-\mu$ is a non-negative integer, which we denote by $n$.

\begin{align*}
\lambda - \mu &= n, \quad \text{for } n = 0, 1, 2, \dots \\
\implies \mu &= \lambda - n. \\
E_n &= -\frac{\mu^2}{2} = -\frac{(\lambda-n)^2}{2}.
\end{align*}

These are the discrete energy eigenvalues corresponding to the bound states of the potential. For these states to exist, the energy must be negative, which is satisfied for $\lambda>n$. Furthermore, for the wavefunction to be normalizable, we require $\lambda>n$. The number of bound states is therefore finite and given by the largest integer $n$ strictly less than $\lambda$. The corresponding eigenfunctions are
\begin{equation}
\phi_n(y) \propto P_\lambda^{\lambda-n}(\tanh(y)).
\end{equation}

For energies $E>0$, the solutions correspond to a continuous spectrum of scattering states and are described by hypergeometric functions \cite{Flugge1999}.

\section{Special Functions}
\label{sec:AppendixB}

This appendix provides the definitions and governing equations for the special functions that appear as solutions to the Sturm-Liouville problem derived in the main text. The functions presented are standard in the literature of mathematical physics and are essential for constructing the eigenfunctions of the Pöschl-Teller Hamiltonian \cite{AbramowitzStegun1964, WhittakerWatson1927}.

The associated Legendre functions, denoted $P_\nu^\mu(z)$, are solutions to the general Legendre differential equation. These functions form the basis for the bound state eigenfunctions of the hyperbolic Pöschl-Teller potential, as detailed in Section \ref{sec:AppendixA}.

The general Legendre differential equation is given by:
\begin{align*}
(1-z^2)\frac{d^2w}{dz^2} - 2z\frac{dw}{dz} + \left[\nu(\nu+1) - \frac{\mu^2}{1-z^2}\right]w &= 0.
\end{align*}

The solutions $w(z)$ are the associated Legendre functions of the first kind, $P_\nu^\mu(z)$, and second kind, $Q_\nu^\mu(z)$. For the bound state problem, physical solutions require square-integrability, selecting the $P_\nu^\mu(z)$ functions.

An integral representation for $\Re(\nu+\mu) > -1$ is given by:
\begin{align*}
P_\nu^\mu(z) &= \frac{(\nu+1)\cdots(\nu+\mu)}{(\nu-1)\cdots(\nu-\mu+1)} \frac{1}{\Gamma(1-\mu)} \int_1^z \frac{(t^2-1)^{-\mu/2}}{(z-t)^{1-\mu}} P_\nu(t) dt.
\end{align*}

In the context of this paper, the argument is $z = \tanh(y)$, and the degrees $\nu, \mu$ are determined by the quantum numbers.

The Gauss hypergeometric function, denoted ${}_2F_1(a,b;c;z)$, is a more general special function that arises as the solution to the hypergeometric differential equation. It describes the continuous spectrum of scattering states for the Pöschl-Teller potential.

The hypergeometric series is defined for $|z| < 1$ by the power series:
\begin{align*}
{}_2F_1(a,b;c;z) &= \sum_{n=0}^\infty \frac{(a)_n (b)_n}{(c)_n} \frac{z^n}{n!}.
\end{align*}

Here, $(q)_n$ is the Pochhammer symbol (or rising factorial), defined as:
\begin{align*}
(q)_n &= q(q+1)\cdots(q+n-1) = \frac{\Gamma(q+n)}{\Gamma(q)}, \quad (q)_0 = 1.
\end{align*}

The function $w(z) = {}_2F_1(a,b;c;z)$ is the principal solution to the hypergeometric differential equation:
\begin{align*}
z(1-z)\frac{d^2w}{dz^2} + [c-(a+b+1)z]\frac{dw}{dz} - abw &= 0.
\end{align*}

The solutions for the scattering states of the Pöschl-Teller potential can be constructed from these functions.

A relevant integral representation (Euler's integral) for $\Re(c) > \Re(b) > 0$ is:
\begin{align*}
{}_2F_1(a,b;c;z) &= \frac{\Gamma(c)}{\Gamma(b)\Gamma(c-b)} \int_0^1 t^{b-1}(1-t)^{c-b-1}(1-tz)^{-a} dt.
\end{align*}

Comprehensive tables of properties, identities, and integral representations for these functions can be found in standard references such as Abramowitz and Stegun \cite{AbramowitzStegun1964} and Gradshteyn and Ryzhik \cite{GradshteynRyzhik2007}.

\end{document}